\documentclass[a4paper]{article}
\usepackage[ansinew]{inputenc}
\usepackage{times}
\usepackage[T1]{fontenc}
\usepackage{graphicx}
\usepackage{geometry}
\usepackage{amssymb}
\usepackage{amsmath}

\usepackage{rotating}

\usepackage{xcolor}
\colorlet{shadecolor}{gray!25}
\usepackage{subfig}
\usepackage{float}
\usepackage{authblk}

\title{Model-based simultaneous inference for multiple subgroups and multiple endpoints}
\author[1]{Charlotte Vogel}
\author[1]{Frank Schaarschmidt}
\author[2]{Christian Ritz}
\author[3]{Franz Koenig}
\author[1]{Ludwig A. Hothorn}

\affil[1]{\small{Institute of Cell Biology,Biostatistics, Leibniz University of Hanover, Herrenhauser Strasse 2, 30419 Hannover, Germany}}
\affil[2]{Department of Nutrition, Exercise and Sports, Faculty of Science, University of Copenhagen, Denmark}
\affil[3]{Institute of Medical Statistics, Medical University of Vienna, Austria}

\begin{document}

\maketitle
\begin{abstract}
\noindent
Various methodological options exist on evaluating differences in both 
subgroups and the overall population. Most desirable is the simultaneous study of 
multiple endpoints in several populations. We investigate a newer method using 
multiple marginal models (mmm) which allows flexible handling of multiple 
endpoints, including continuous, binary or time-to-event data. 
This paper explores the performance of mmm in contrast to the standard 
Bonferroni approach via simulation. Mainly these methods are compared on the 
basis of their familywise error rate and power under different scenarios, 
varying in sample size and standard deviation. Additionally, it is shown that 
the method can deal with overlapping subgroup definitions and different 
combinations of endpoints may be assumed. The reanalysis of a clinical example 
shows a practical application.
\end{abstract}

\section{Introduction}

The analysis of subgroups is challenging in randomized phase III clinical trials both for 
a priori defined and post hoc selected subgroups. What is particularly challenging is the exact analysis and the appropriate interpretation between global and subgroup-specific consideration \cite{Dane2019}. Particularly, observed effect sizes 
may differ between subgroups and therefore the interpretation of results can become ambitiously. 
What does it mean when a subgroup shows a larger effect than the overall population? For instance Mok et. al. \cite{Mok2009} published a study, where the $0.64$ upper confidence limit of the hazard ratio of progression-free survival in the EGFR-mutation positive subgroup reveals a substantially larger benefit compared with $0.85$ limit in the overall population after treatment of patients with pulmonary adenocarcinoma with Gefitinib compared with Carboplatin-Paclitaxel. A guideline was released by the EMA \cite{EMA2019} focusing on 
the exploratory investigation of subgroups where the treatment effects are in line with the overall population or show different if not contradictory results. 
Principally, it supports methods that take care of multiplicity and are sensitive to treatment effects, so that a further assessment of efficacy is possible. Accordingly, our focus lies upon the stringent requirement, that \textit{"It is highly unlikely that claims based on subgroup analyses would be accepted in the absence of a significant effect in the overall study population"} \cite{CPMP1999}.\\ 
Up to now controversies exist on several aspects, such as the inherently reduced 
power for subpopulations or the need of an interaction test including the option 
to test for clinical relevance in components next to the global null hypothesis. 
To alleviate the fact that interaction tests lack power, the choice of subgroups must be sufficiently considered even more.
Often it can be assumed that already a preferred subgroup of interest exists. In the research of biomarkers, it is common for a subgroup to be composed only with regard to the positive expression of a biomarker ($B^+$) whereby the other set of observations merges into a 
heterogeneous population of other cases. The latter are therefore not of equal interest 
and it would be logically inconsistent to test them as a remaining subgroup \cite{Glimm}.
This scenario is extended when there is no or minimal effect on the entire
population but a beneficial effect can be demonstrated in the targeted subgroup. Using a fallback procedure that also accounts for the correlation between test statistics, both the hypotheses of the overall study population and the subgroup may be tested once a pre-specified degree of consistency is met \cite{Alosh2009}.
Another scenario has to be considered when licensing for the total population requires
additionally to its own significance that the treatment effect in the
complementary subgroup is above a threshold. Otherwise, the claim is made for 
the targeted subgroup only (consistency constraint) \cite{Alosh2013}.\\
Alternatively, without restricting the alternative by any a priori assumptions 
we propose an union-intersection test (UIT) principle that makes it possible to simultaneously test for a treatment effect in any combination of populations to be studied, e.g. at least one subgroup or total population.
The global null hypothesis of neither effect in each population is contrasted against the alternatives of at least one - any one - significant population or any patterns up to all significant populations.
Clearly, that is a selected conservative approach 
whereas its conservativeness is reduced by taking the correlations of the test statistics into account. 
The advantage is the availability for claims of any population patterns 
(e.g. global and targeted subgroup) and, instead of adjusted p-values or 
alpha-propagation rules \cite{Millen2012}, the use of simultaneous confidence 
limits for appropriately chosen effect sizes (see the recommendation in ICH E9 \cite{FDA1998}).\\
Recently, a case study \cite{Bretz2011} was presented in which an oncology 
trial explores two primary endpoints, progression-free survival after 2.5 years 
and overall survival after 4 years, and their examination in three populations. 
Yet it is not uncommon to select multiple endpoints and further extend this 
interest in more than one population. Consequently, we discuss the UITs for both some subgroups and multiple primary 
endpoints without any restriction in importance. Specifically, a different scaling is allowed for different endpoints.\\
This paper will compare the combination of marginal linear models 
\texttt{mmm} \cite{Ritz2012} by using the multivariate normal or 
multivariate t distribution with a different choice of degrees of freedom 
in contrast to a simple Bonferroni approach or multiple contrast test
if the correlation structure between test statistics is known. 
We investigate whether the methods control the familywise error rate in 
small sample situations and provide advantages in power. 
It is shown that \texttt{mmm} can deal with overlapping subgroup definitions and 
offers an easy handling of multiple endpoints, also with different distributions 
or different scales, such as continuous, proportions or time-to-event data. 
Finally accounting for correlation leads to a benefit in power compared to 
Bonferroni.\\
Section 3 provides a statistical framework for multiple pairwise comparisons 
used in this paper. Simulation results for familywise error rate and power are 
examined in Section 4. An application of the \texttt{mmm} method in \texttt{R} is 
illustrated in Section 5 by one clinical example. Section 6 concludes the 
article with a discussion and a few comments.


\section{A Case Study: the AVEROES trial}
In a randomized two-arm trial Apixaban, a novel factor Xa inhibitor was compared 
with aspirin to reduce the risk of stroke or systemic embolism in patients
with atrial fibrillation for the primary efficacy outcome of stroke (ischaemic, 
haemorrhagic, or unspecified stroke). A predefined subgroup analysis was 
performed whether patients with previous stroke or transient ischaemic attack (S1) 
would show a greater benefit from Apixaban than patients with no previous stroke 
or TIA (S2) \cite{Diener2012}. In Table \ref{table:rates} 
the event data in patients with and without a history of stroke or TIA is given 
in a dichotomous structure counting the numbers of stroke or systemic embolism 
and no such an event for the total population, both subgroups and the subordinated
endpoints (Ischemic or unspecified stroke, Hemorrhagic stroke, and stroke in general).
 
\begin{table}[ht]
\centering
\begin{tabular}{r|r||rr|rr|rr}
  \hline
          &        		& \multicolumn{2}{c|}{Global} & \multicolumn{2}{c|}{S1} & \multicolumn{2}{c}{S2} \\ 
  \hline
 Treatment&Endpoint& no Event & Event &no Event & Event & no Event & Event \\ 
  \hline
	Apixaban & Ischaemic &2764 &  43 & 381 &   9 & 2383 &  34 \\ 
	Aspirin & 					& 2692 &  97 & 347 &  27 & 2345 &  70 \\ 
	Apixaban & Hemorrhag & 2801 &   6 & 389 &   1 & 2412 &   5 \\ 
	Aspirin & 					& 2780 &   9 & 370 &   4 & 2410 &   5 \\ 
	Apixaban & Stroke		&2758 &  49 & 380 &  10 & 2378 &  39 \\ 
	Aspirin & 					&2684 & 105 & 344 &  30 & 2340 &  75 \\ 

   \hline
\end{tabular}
\caption{\textbf{Selected effects of apixaban on efficacy outcomes in the full population and in patients with and without history of stroke or TIA (transient ischaemic attack), taken from \cite[p. 228]{Diener2012}.}}
\label{table:rates}
\end{table}


\section{Methods}

We consider a two-way ANOVA model
\begin{equation}
	Y_{ijk} = \mu_{ij} + \epsilon_{ijk}, \qquad 1 \leq i \leq m, \quad 1 \leq j \leq l, \quad 1 \leq k \leq n_{ij},
	\label{eq:model}
\end{equation}
where $y_{ijk}$ denotes the $k$-th observation in subgroup $j$ of treatment group $i$, $\mu_{ij}$ is the mean in subgroup $j$ of treatment group $i$ and $\epsilon_{ijk}$ are random independent and normally distributed errors. The treatment factor has $m$ levels with all groups $i$ subdivided into $l$ subgroups in which subgroup $j$ contains $n_{ij}$ observations. It is assumed that with $\alpha_i = \mu_{i.} - \mu$, $\beta_j = \mu_{.j} - \mu$ and $(\alpha \beta)_{ij} = \mu_{ij} -  (\mu + \alpha_i + \beta_j)$ the model can be rewritten as
\begin{equation}
	Y_{ijk} = \mu + \alpha_i + \beta_j + (\alpha \beta)_{ij} + \epsilon_{ijk},
\end{equation}
where the parameter $\mu$ denotes the grand mean, $\alpha_i$ corresponds to the main treatment effect in group $i$, $\beta_j$ to the effect in subgroup $j$ and $(\alpha \beta)_{ij}$ to the joint effect of the $i$-th level of the treatment factor and the $j$-th level of the subgroup factor.
We are then interested in the superiority of e.g. a new treatment in each subgroup and the total population compared to placebo which can be expressed for the standard case of $m=2$ treatment groups and $l=2$ subgroups by the null hypotheses
\begin{enumerate}
	\item 	$H_{total}: \alpha_1 - \alpha_2 = 0$ (no difference in total population)
	\item 	$H_{target}: \mu_{11} - \mu_{21} = 0$ (no difference in subgroup $j=1$, denoted as targeted)
	\item 	$H_{compl}: \mu_{12} - \mu_{22} = 0$ (no difference in subgroup $j=2$, denoted as complementary).
\end{enumerate}


\subsection{Multiple Marginal Models (mmm)}
A flexible approach has been introduced by Pipper et al. \cite{Ritz2012} in which a set of statistical inferences concerning the same sample can be assessed by a formulation of multiple marginal models. All models can be evaluated simultaneously by estimating the correlation between the test statistics using a score decomposition and hence no explicit formulation of the correlation is required.\\
For each model fit maximum likelihood estimators have asymptotic representations based on standardized score functions for observations \cite[Theorem 5.21]{vdV00}. These asymptotic representations may be combined into a multivariate asymptotic representation by stacking over all models (stacking is not destroying independence). Convergence in distribution of the corresponding stacked parameter estimates is ensured by the multivariate central limit theorem. Moreover, a consistent estimator of $\Sigma$ may be obtained as the empirical variance-covariance of the stacked standardized score functions. This asymptotic result implies the following multivariate normal approximation to the family-wise error rate for test statistics $Z_r$ (significance level $\alpha$):
\begin{align*}
	P(max_{r=1,\ldots,R} |Z_r| > z_{1-\alpha/2}) &\to 1-\int_{-z_{1-\alpha/2}}^{z_{1-\alpha/2}} \ldots \int_{-z_{1-\alpha/2}}^{z_{1-\alpha/2}} \phi(s, 0, C) ds\\ 
	&= f_C(\alpha)
\end{align*}
where $z_{1-\alpha/2}$ is the $1-\alpha/2$ percentile in $N(0,1)$, $\phi$ the $R$-dim. multivariate normal density, and $C$ \linebreak the variance-covariance matrix of $(Z_1, \ldots, Z_R)$. A consistent estimator of $C$ is given by \linebreak $\hat{C} = diag(\hat{\Sigma})^{-1/2} \hat{\Sigma} diag(\hat{\Sigma})^{-1/2}$. We refer to \cite{Ritz2012} for additional details (including proofs of weak and strong control of the family-wise error rate). The resulting multiplicity adjustment is always less conservative than the Bonferroni adjustment (the larger the correlation between test statistics the larger the gain).\\
Various statistical models can be used as marginal models. For example in the recently introduced event of an overall model with all observations included and a number of smaller models for subsets of data i.e. targeted and complementary subgroup where observations which do not belong to the subgroup are set to missing. Basically in a case of two treatment arms $(l + 1)$ univariate linear models are fitted, one for each subgroup comparison plus one for the inference decision in the global population.\\
For the standard example of one treatment group ($trt$) and one control group ($ctrl$) i.e. $m=2$ treatment groups with $l=2$ subgroups the set of three fitted models is: 1) $trt$ vs. $ctrl$ with observations of the second subgroup set to missing, 2) $trt$ vs. $ctrl$ with observations of the first subgroup set to missing and 3) $trt$ vs. $ctrl$ for the main effect with all observations available. P-values and confidence intervals can be then adjusted using a reference distribution based on the estimated correlation matrix of those models.\\
In \texttt{R} the functions \texttt{mmm()} and \texttt{glht()} are implemented in the \texttt{R} package \texttt{multcomp} \cite{multcomp} for the calculation of the correlation matrix and simultaneous testing of hypotheses respectively. In the basic formulation of \texttt{glht()}, when no degrees of freedom are stated, results rely on a multivariate normal distribution. Otherwise, degrees of freedom may be specified as an additional \texttt{df} argument to \texttt{glht()} and the multivariate \textit{t} distribution is used for the evaluation.

   
\noindent
A complicating factor is added when the various patient subpopulations arise from overlapping subgroup definitions which are not disjoint. This is obviously the case if several factors of interest exist and the association is expressed dichotomously in yes/no, e.g. in biomarker studies or multiple nominal factors are combined to one subgroup interpretation, for instance $S1:$ node-negative, no chemotherapy and $S2:$ any nodal status, no chemotherapy \cite{SR13}. Even more than one endpoint might be important in a clinical study. Rather than defining one primary endpoint, several models can be established with respect to different endpoints. Both complications of overlapping subgroups and multiple endpoints, even a combination can be incorporated in individual model definitions in \texttt{mmm}.\\
Marginal models may not only be applied to linear models with Gaussian error terms. It also features generalized linear models that have a non-normal error distribution like binomial, multinomial, poisson or negative binomial distribution covering virtually all statistical circumstances. Furthermore \texttt{mmm} allows for covariate adjustment with different covariance structure per endpoint or subgroup definition. 

\subsection{Simulation}

\subsubsection{Two treatments and subgroups in the general linear model}
For a start, we consider the simplest statistical setting of two equally sized treatment groups ($m=2$) and two subgroups ($l=2$) for which we assume normality of the residuals and homogeneity of variance. To assess potential differences in treatment the following test procedures are investigated in this paper. The corresponding abbreviation used for the figures is named in italics.

\textbf{No multiplicity adjustment (\textit{noadjust}).} This procedure of no adjustment in which the situation of multiple comparisons is completely ignored is only included for completeness and does not control the familywise error rate by definition. The p-values are calculated from a set of three marginal tests, e.g. three t-tests, where one is for the main group difference ($trt$ vs. $ctrl$) and one for each subgroup comparisons ($trt_{target}$ vs. $ctrl_{target}$ and $trt_{compl}$ vs. $ctrl_{compl}$). These p-values will be each compared to a comparison-wise significance level of 0.05 (5\%).
   
\textbf{Bonferroni (\textit{bonferroni}).} Bonferroni correction is a fairly simple and still common way of handling multiplicity in clinical studies and aims to control the familywise error rate although it is conservative. The p-values derived from the three t-tests mentioned above (i.e. the nonadjust approach) will be compared at $\alpha = 0.05 / 3$. Correlation of the test statistics is completely ignored in this case. 
    
\textbf{Cell-means model (\textit{cellmeans}).} In this approach a fitted response model (\ref{eq:model}) with parameters $(\mu_{11},\mu_{21},\mu_{12},\mu_{22})$, which are called the cell means, and a contrast matrix $C$ 
	\[C = \begin{pmatrix}
      -1 & 1 & 0 & 0 \\
      0 & 0 & -1 & 1 \\
      - \frac{n_{11}}{(n_{11}+n_{12})} & \frac{n_{21}}{(n_{21}+n_{22})} & - \frac{n_{12}}{(n_{11}+n_{12})} & \frac{n_{22}}{(n_{21}+n_{22})}
     \end{pmatrix}
\]
can be used to simultaneously test treatment effects within subgroups (first and second contrast) and the main effect of treatment across subgroups (third contrast). Note that the elements of $\mu$ or columns of $C$ respectively are primarily ordered according to the subgroup, and within that factor according to treatment. For the third contrast each mean of the treatment group is estimated by the mean of the observations within the subgroups weighted according to their proportion of the treatment group. So $\frac{n_{11}}{(n_{11}+n_{12})} \mu_{11} + \frac{n_{12}}{(n_{11}+n_{12})} \mu_{12}$ represents the weighted arithmetic mean of the means in subgroups 1 and 2 in the control group and the weights $\{\frac{n_{11}}{(n_{11}+n_{12})}, \frac{n_{12}}{(n_{11}+n_{12})}\}$ are their proportional sample sizes in the control. All p-values will be adjusted for multiple comparisons using the known correlation matrix and the underlying trivariate $t$ distribution \cite{GenzBretz2009}. In \texttt{R} this is made easy by the access of \texttt{multcomp} \cite{multcomp} on the \texttt{mvtnorm} package (\cite{mvtnormMan}). 


\textbf{Multiple Marginal Models (\textit{mmm}).} 
The approach of Multiple Marginal Models (\texttt{mmm}) calculates the correlation between all test statistics using a score decomposition \cite{Ritz2012} and hence no explicit formulation of the correlation is required.
P-values will be adjusted using a reference distribution, based on the estimated correlation matrix. In the basic formulation, when no degrees of freedom are given, results rely on a multivariate normal distribution (\textbf{\textit{mmm}}). If degrees of freedom are specified, for example, the smallest degree of freedom of all models (\textbf{\textit{mmm.dfmin}}), the largest degree of freedom of all models (\textbf{\textit{mmm.dfmax}}) or model specific degrees of freedom (\textbf{\textit{mmm.dfind}}) multivariate t distribution is used for the evaluation. This approach is implemented via \texttt{mmm()} in the \texttt{R} package \texttt{multcomp} and can be customized by specifying \texttt{df} as an additional argument to \texttt{glht()}. 

\subsubsection{Simulation structure}

In a simulation study, we addressed the individual impact of different distribution parameters on the familywise error rate and power for all considered methods. 
The sample size varied from small to large with a total number of observations $N = 20, 50, 100, 500$ equally distributed to the treatment groups. Since no heteroscedasticity was assumed, equal standard deviations were chosen for all subgroups varying in different scenarios as $\sigma = (2, 5, 10)$. Moreover, we studied unbalanced subgroup sizes with various proportions of the targeted subgroup ranging from $0.5$ to $0.8$ increased in constants of $0.1$. The overall mean was set to $\mu = 0$ and an effect of $\delta = 0, \dots, 10$ was added to each observation of the first subgroup of the treatment group.\\
Multiple endpoints were generated from a bivariate normal distribution with the same assumptions from above and different correlation parameters $\rho = 0.2, \dots, 0.8$.\\
To investigate the handling of overlapping subgroups a second subgroup definition was introduced with same the expectation of subgroups ratios.\\
For each scenario, a number of $10 000$ datasets was generated and all comparisons of the treatment effect in the subgroups and in the overall population were tested according to each method. Since all comparisons were false for $\delta = 0$ the proportion of datasets, in which at least one significant difference was detected for any comparisons ("any") or for the targeted subgroup or the whole population ("targeted or total"), represents the familywise error rate of the method. The power was estimated as the proportion of cases with at least one hypothesis in any comparison correctly rejected. 
All simulations were performed in \texttt{R}, version 3.2.1 \cite{Rprogram}. 



\section{Simulation Results}

		\begin{figure}[htb]
		\includegraphics[width=16cm, height=9cm]{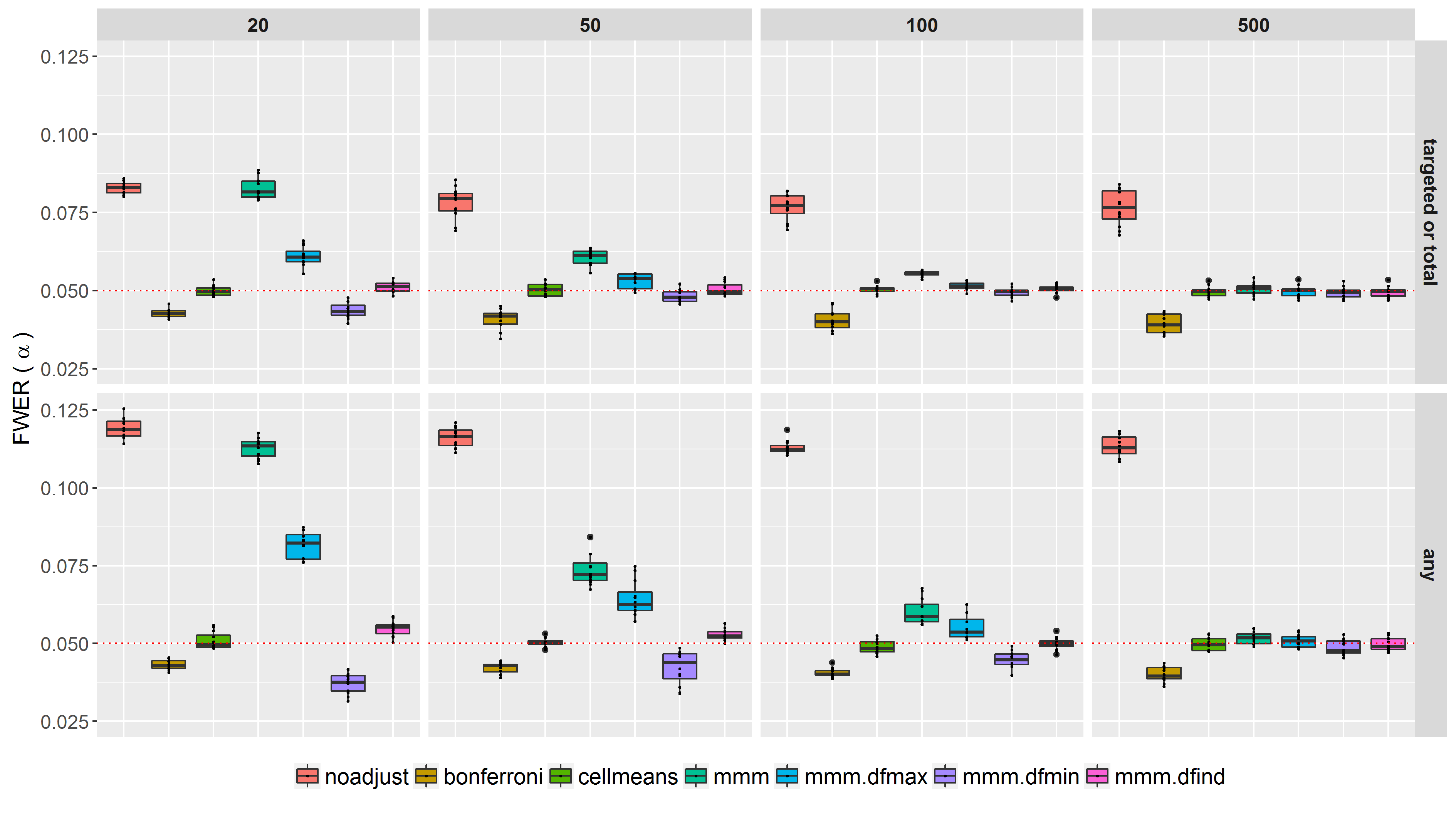}%
		\caption{
		\textbf{Simulated familywise error rate for simultaneous test procedures 
		when testing for different sets of hypotheses.} 
		Comparison of the estimated familywise error rate among two sets of hypotheses: 
		first, that there is a difference in the whole population ($H_{total}$) and 
		the targeted subgroup ($H_{target}$) labelled by "targeted or total" in the 
		top row and second, that there is a difference in the whole population and 
		in both of the subgroups labeled by "any" in the bottom row. All rates are 
		assessed for different sample sizes and averaged over unbalanced subgroup 
		sizes and varying standard deviations under homoscedasticity. The total 
		sample size $N$ was equally distributed to the two treatment groups. 
		The a priori chosen alpha level of $0.05$ is represented by the horizontal 
		red dotted line.}
		\label{fig:alpha}
		\end{figure}

\subsection{Type I error in the simultaneous analysis of multiple subgroups}

To determine the extent to which the approaches identify an effect between the treatments, it should first of all not be important whether this difference is present in the total population or in one of the subgroups. In case that just one subgroup is of interest, this subset is referred to as targeted. This means that the null hypothesis is tested in the total population and in the targeted subpopulation only and at least one of these hypotheses is rejected to consider significance. Otherwise all subgroups are included in the definition of the hypotheses and are examined at the same time.
Results of the simulation study will be compared for all procedures, which have been described in the methods section in detail, with respect to the preservation of the familywise error rate given moderate or large sample sizes. Continuous endpoints are examined first. An overlap of subgroups is incorporated next and multiple endpoints are examined in addition. The power is only shown for those methods that control for the probability of making a type I error.

\subsubsection{Concerning one targeted subgroup.}

Figure \ref{fig:alpha} illustrates the estimated familywise error rates of the simultaneous test procedures averaged across different parameter settings like varying standard deviations and subgroup populations. In the upper row, the type I error is shown in the case of testing for significant differences in the whole population and the targeted subgroup. The simulation results are shown depending on the total number of observations $N$ and differ by the test procedure. In a situation of multiple comparisons, no adjustment shows harsh violations of the familywise error rate. Taking into account the multiplicity issue all \texttt{mmm} methods exceed the 0.05 level in situations with small treatment groups except for \texttt{mmm.dfmin} which specifies the smallest degree of freedom. Clearly, it is shown that \texttt{mmm} is an asymptotic procedure and best suitable for large sample sizes. Its performance can be improved by using a multivariate \textit{t} reference distribution. For example \texttt{mmm.dfind} shows a familywise error rate near the 0.05 level for all sample sizes and is less conservative than the \texttt{Bonferroni} adjustment. Likewise the \texttt{cellmeans} model controls the FWER to full exploitation. 
	
\subsubsection{Concerning any subgroup.}

In case that no subgroup is preferred the estimated familywise error rates are illustrated in the second row of Figure \ref{fig:alpha} (any). Since all comparisons must be taken into account the results are slightly inflated than these for the first simulation study. From a sample size of $250$ subjects per group ($N=500$) onwards the estimated type I error levels off at $\alpha = 0.05$ for all \texttt{mmm} adjustment procedures based on correlation estimation. For lower sample sizes the methods show heterogeneous results depending on the before specified degree of freedom. \texttt{mmm.dfind} slightly exceeds the type I error. Again the \texttt{cellmeans} model controls the FWER.

\subsection{Simultaneous analysis of several overlapping subgroup definitions}

\begin{figure}[htb]
		\includegraphics[width=16cm, height=9cm]{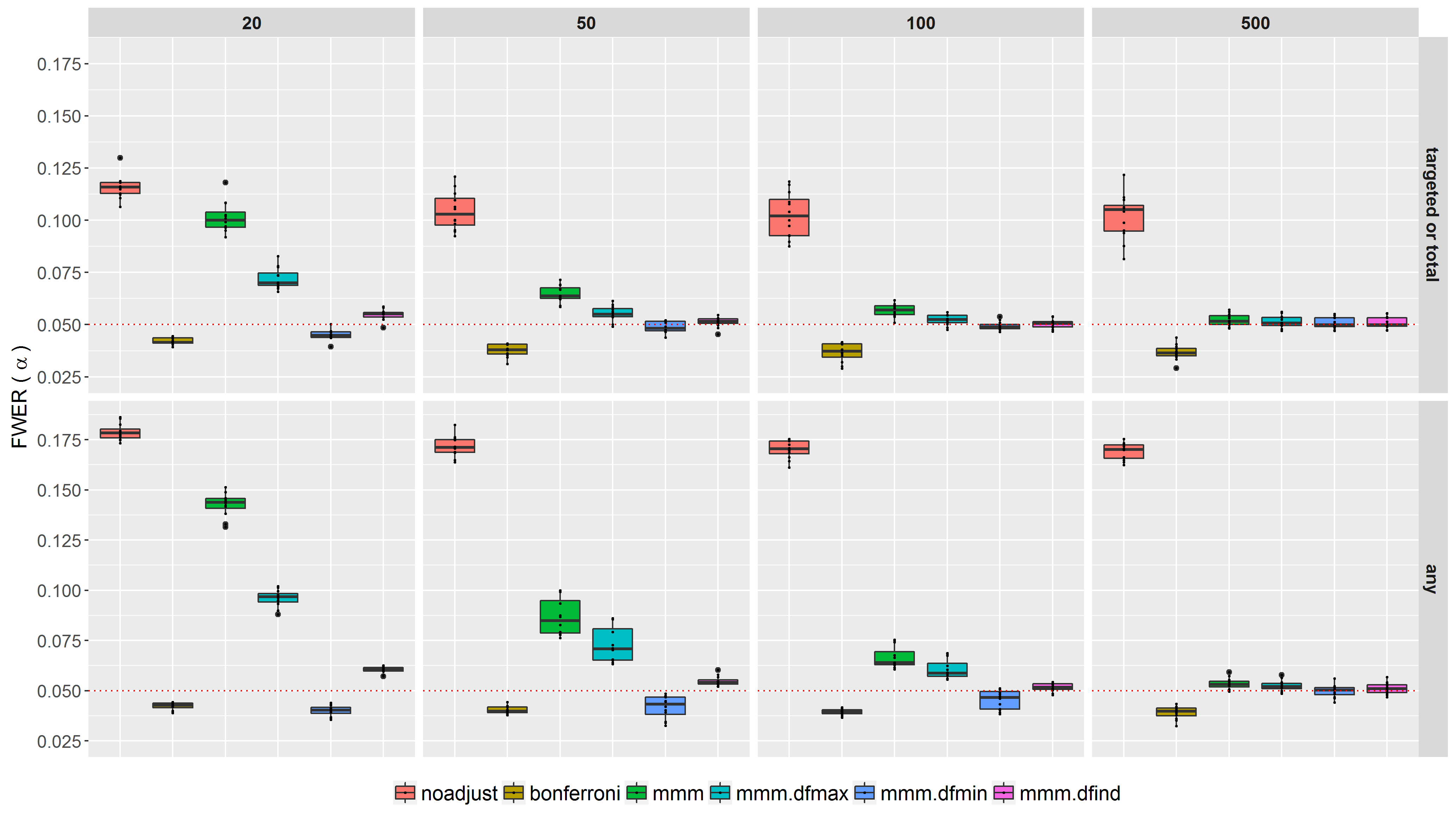}%
		\caption{
			\textbf{Simulated familywise error rate for simultaneous test procedures 
			for two overlapping subgroups.} 
			Comparison of the estimated familywise error rate in case of two 
			overlapping subgroup definitions but one primary endpoint for both 
			families of hypotheses, first that there is a difference in the whole 
			population and the targeted subgroup and second, that there is a 
			difference in the whole population and in both of the subgroups. The rates 
			are calculated for different sample sizes and averaged over unbalanced 
			subgroup sizes and varying standard deviations under homoscedasticity. 
			The total sample size $N$ was equally distributed to the two treatment 
			groups. The a priori chosen alpha level of $0.05$ is represented by the 
			horizontal red dotted line.}
		\label{fig:overlap}
		\end{figure}

On condition of overlapping subgroups the \texttt{cellmeans} method is not considered any further for it cannot deal with more than one single simultaneous model. Therefore Figure \ref{fig:overlap} illustrates the methods of no adjustment, numerous t-tests with Bonferroni correction and \texttt{mmm} only. In all cases, the estimated type I error is above the chosen alpha level of $0.05$ when using no adjustment. Applying Bonferroni to a set of t-tests is highly conservative while both methods \texttt{mmm.dfmin} and \texttt{mmm.dfind} achieve rates below $\alpha = 0.05$ already for a total sample size of $N=50$. The same is true for the simulation in the whole set of hypotheses.

\subsection{Simultaneous analysis of several endpoints}
	
\begin{figure}[htb]
		\includegraphics[width=16cm, height=9cm]{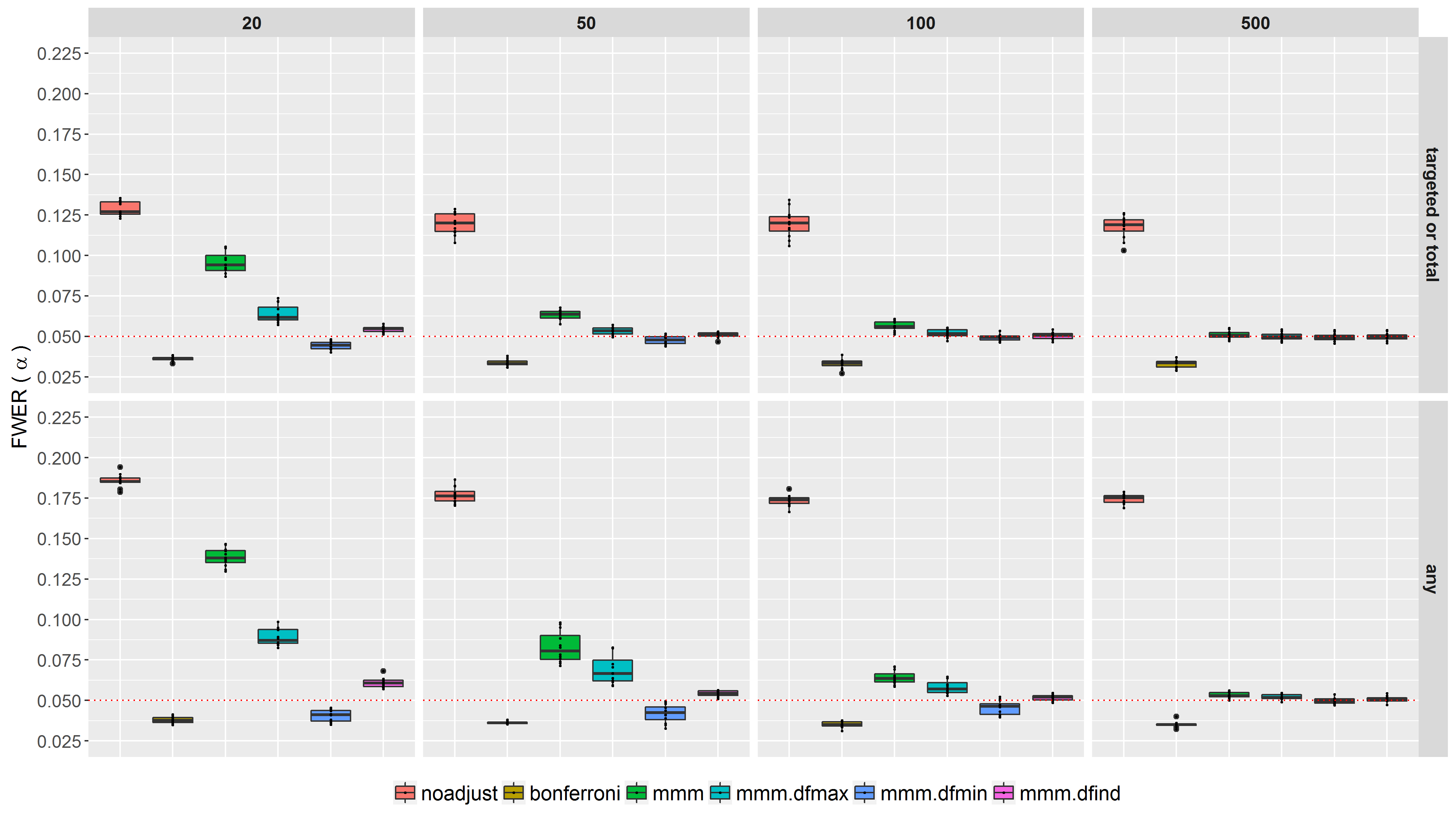}%
		\caption{
			\textbf{Simulated familywise error rate for simultaneous test procedures 
			when testing for a set of hypotheses regarding two continuous endpoints.} 
			Comparison of the estimated familywise error rate first among the two 
			hypotheses that there is a difference in the whole population and the 
			targeted subgroup and second among all hypotheses that there is a 
			difference in the whole population and in both of the subgroups. 
			The rates are calculated regarding two continuous, highly correlated 
			endpoints ($\rho = 0.8$) for different sample sizes and 
			averaged over unbalanced subgroup sizes and varying standard deviations 
			under homoscedasticity. The total sample size $N$ was equally distributed 
			to the two treatment groups. 
			The a-priori chosen alpha level of $\alpha = 0.05$ is represented by the 
			horizontal red dotted line.}
		\label{fig:2ep}
		\end{figure}

Often even more than one endpoint is of major interest in clinical trials. In Figure \ref{fig:2ep} two continuous, highly correlated endpoints ($\rho = 0.8$) endpoints are considered and tested in each of the individual populations. Again \texttt{mmm} provides an easy handling of multiple endpoints and accounts for multiplicity among correlated continuous endpoints. Simulations show similar results: Still the FWER is controlled by \texttt{Bonferroni} for two endpoints but stays conservative in comparison to other multiplicity approaches. With increasing sample sizes the mean estimated familywise error rate of all \texttt{mmm} methods, \texttt{mmm.dfind} in particular, rapidly reduces to the predetermined alpha level.

\subsection{Power of \texttt{mmm}}


Having demonstrated which methods provide adequate protection against a familywise type I error it is now examined whether these methods differ in terms of statistical power. Therefore this section focuses on relevant methods only, namely \texttt{Bonferroni}, \texttt{cellmeans} and \texttt{mmm.dfind} for multiple subgroups according to one endpoint and \texttt{mmm.dfind} compared with \texttt{Bonferroni} for the analysis of the second simulation study regarding two correlated endpoints.

	\begin{figure}[htb]
	\includegraphics[width=16cm, height=9cm]{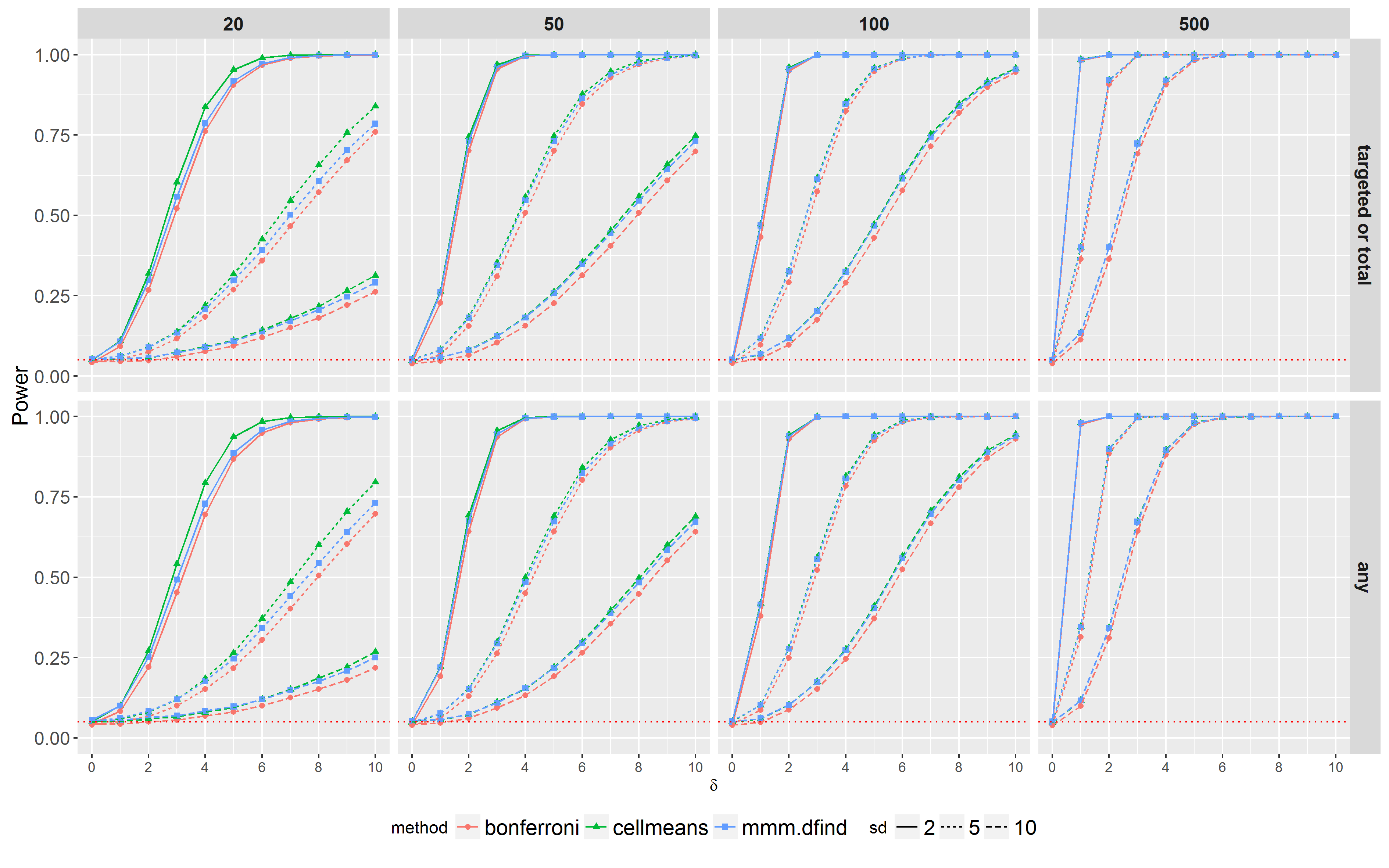}
	\caption{
		\textbf{Simulated power for simultaneous test procedures when testing for 
		hypotheses regarding one continuous endpoint.} 
		Comparison of the estimated power among two sets of hypotheses: first, that 
		there is a difference in the whole population ($H_{total}$) and the targeted 
		subgroup ($H_{target}$) labelled by "targeted or total" in the top row and 
		second, that there is a difference in the whole population and in both of 
		the subgroups labelled by "any" in the bottom row. All rates are assessed 
		for different sample sizes and averaged over unbalanced subgroup sizes. 
		The different line types of the power curves relate to different standard 
		deviations, which are assumed homogeneously in the treatments groups, 
		starting with sd = 2 at the top, sd = 5 in the middle and sd = 10 below. 
		The total sample size $N$ was equally distributed to the two treatment groups. 
		The a-priori chosen alpha level of $ 0.05$ is represented by the horizontal 
		red dotted line.}
	\label{fig:power_normal}
	\end{figure}	

\begin{figure}[htb]
	\includegraphics[width=16cm, height=9cm]{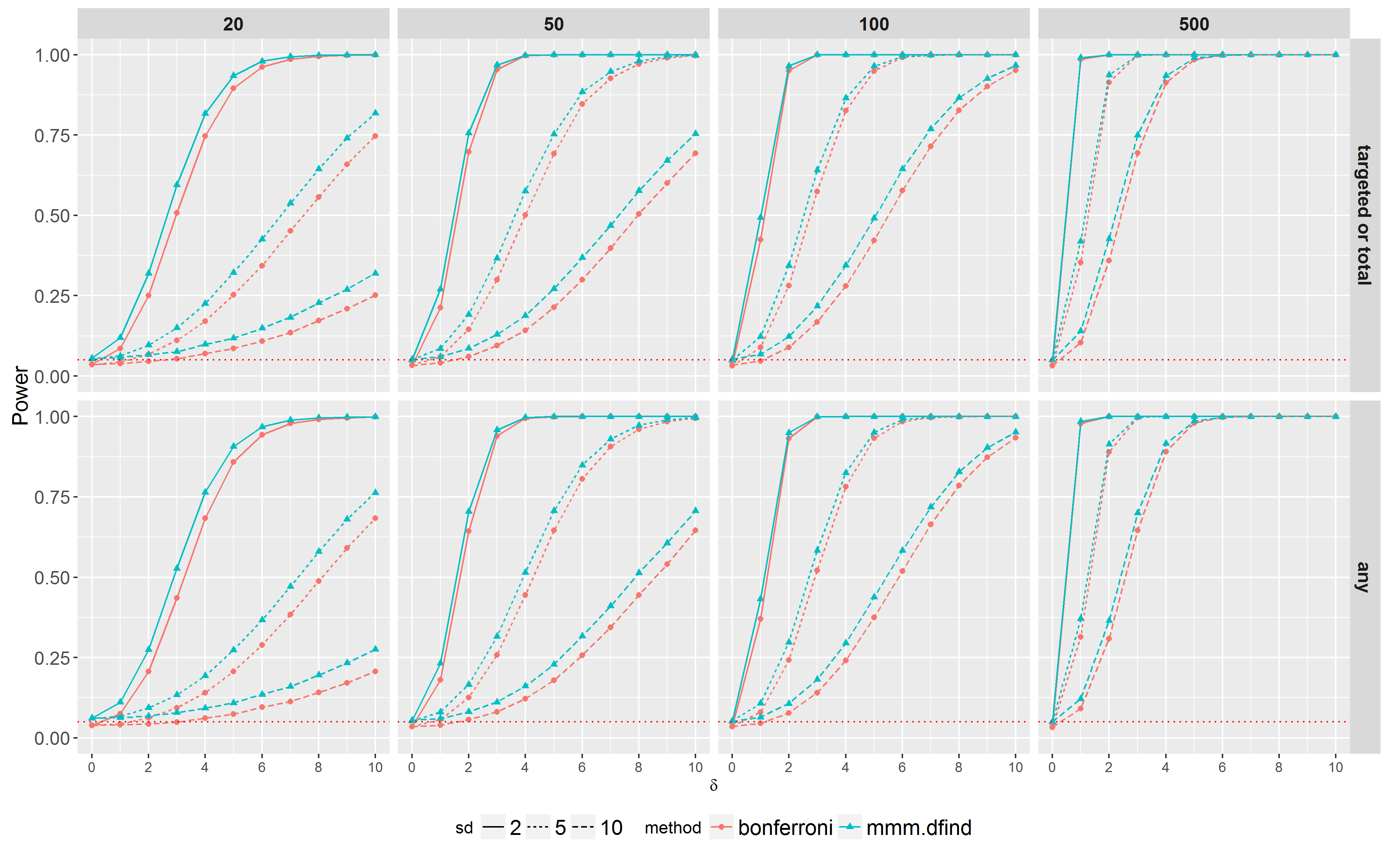}%
	\caption{
		\textbf{Simulated power for simultaneous test procedures when testing for 
		hypotheses regarding two continuous correlated endpoints.} 
		Comparison of the estimated power among two sets of hypotheses: first, that 
		there is a difference in the whole population ($H_{total}$) and the targeted 
		subgroup ($H_{target}$) labelled by "targeted or total" in the top row and 
		second, that there is a difference in the whole population and in both of 
		the subgroups labelled by "any" in the bottom row. Endpoints were simulated 
		with a correlation of $\rho = 0.8$.	All rates are assessed for different 
		sample sizes and averaged over unbalanced subgroup sizes. The different 
		line types of the power curves relate to different standard deviations, 
		which are assumed homogeneously in the treatments groups, starting with 
		sd = 2 at the top, sd = 5 in the middle and sd = 10 below. 
		The total sample size $N$ was equally distributed to the two treatment groups. 
		The a priori chosen alpha level of $0.05$ is represented by the horizontal 
		red dotted line.}
	\label{fig:power_2ep}
	\end{figure}

\noindent
The probabilities to discover a false hypothesis in the primary scenario of one single endpoint regarding multiple non-overlapping subgroups are illustrated in Figure \ref{fig:power_normal} for different parameter settings like varying standard deviations and sample sizes averaged across different subgroup populations when testing for significant differences in the whole population or the targeted subgroup or any population. For larger sample sizes, the power of all methods with multiplicity adjustment is as expected. Major differences in power occur, the smaller the sample size is. For example, the gain of power for a total sample size of $N = 50$ is 5.47\% when one chooses \texttt{mmm.dfind} compared to the combination of several \textit{t}-tests with Bonferroni adjustment ($sd = 10$). Even bigger is the advantage when using the \texttt{cellmeans} model instead, e.g. at a sample size of $N = 20$. Here the benefit reaches up to 13.8\% ($sd = 2$) as a gain in power. At a sample size of $N = 50$, the numeric gain is still more than $6\%$. Figure \ref{fig:power_2ep} represents a similar analysis for the case of multiple endpoints and multiple subgroups. Especially when a strong correlation between endpoints exists the new method \texttt{mmm.dfind} shows a substantial benefit in power. Thus, the power curves of highly correlated endpoints ($\rho = 0.80$) are displayed. In these a power advantage up to $8.35\%$ ($sd = 5$) and $8.50\%$ ($sd = 10$) can be retrieved, e.g. for a sample size of $N = 50$.

\section{Case Study Analysis}
\begin{figure}[htb]
	\includegraphics[width=14cm, height=7cm]{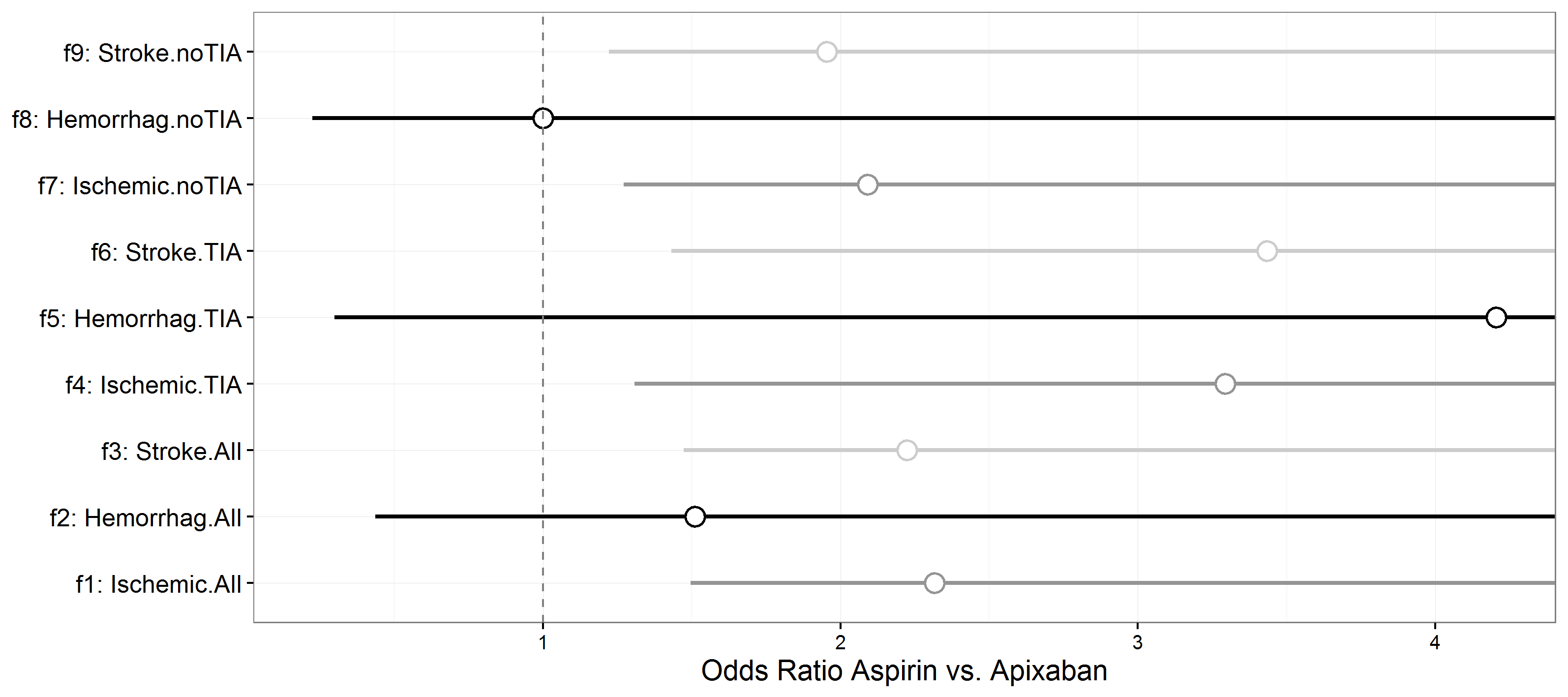}%
	\caption{\textbf{One-sided simultaneous confidence intervals for the Odds 
	Ratio of the AVERROES trial.} Confidence intervals are based on the approach 
	of marginal models taking correlations into account.}
	\label{fig:ci_plot}
	\end{figure}

The treatment effect of each endpoint comparing Apixaban versus Aspirin as the reference group and their 95\% confidence intervals are shown in Figure \ref{fig:ci_plot}. Although Apixaban reduced the risk of stroke and ischaemic or unspecified stroke in the full population and in both subgroups, no further benefit can be demonstrated for the efficacy endpoint of haemorrhagic stroke. The subgroup analysis of hemorrhagic stroke in patients with previous stroke or TIA has a point estimate of $4.21$, and its lower bound of the one-sided confidence interval is at 0.3 (p = 0.39). The results in the total population show a point estimate of $1.51$ with a confidence interval that ranges from
$[ 0.44, \text{Inf} ]$ (p = 0.65), indicating that the events for patients receiving Apixaban are not significantly lower than for patients receiving Aspirin.


\section{Discussion}
\label{sec:Discussion}

Randomized clinical trials should be interpreted by appropriately chosen effect sizes and their simultaneous confidence intervals, whereas p-values should be avoided \cite{FDA1998}. Therefore we propose a UIT-approach allowing all possible patterns of alternatives for total, targeted and complementary subgroups. The cost of such multiple testing is conservativeness whereas it is reduced (with respect to Bonferroni) by taking the correlations between the test statistics into account.\\
Using the \texttt{mmm} approach marginal models can be formulated as generalized linear mixed-effects models allowing the analysis of various primary endpoints occurring in RCTs, including multiple primary endpoints - even different scaled \cite{Jensen2018}. It controls the familywise error rate in simulations with a total sample size of greater than or equal to 500. For smaller sample sizes this method is recommended with an extra declaration of model specific degrees of freedom (\texttt{mmm.dfind}).\\
Extensions to multi-armed clinical trials are possible, e.g. using Dunnett procedure for comparing various treatments against placebo. Of course, no increasing power can be achieved with respect to cell-means model (only compared for non-overlapping subgroups). On the other hand, our new technique shows a remarkable power increase compared to the standard \texttt{Bonferroni} adjustment. The same applies to the case of overlapping subgroup definitions. The performance of the marginal models was generally superior in terms of exploitation of alpha and power. By means of the function "mmm" in the R package multcomp this approach is easily available, even for several types of (multiple) primary endpoints. Also the use of Cox models is possible but has not been shown here \cite{Lin2016}.\\
While the detailed simulations of this paper demonstrate a wide range of application possibilities in settings with multiplicity issues, further investigations may be constructive.
Especially interesting is this novel approach for cases where the explicit formulation of the covariance matrix is difficult or not available. We currently consider unbalanced designs with serious heterogeneity for further research on that topic.

 
\newpage
\appendix
\section{Appendix}
\subsection{Data structure and program code}

For evaluation the dataset has to be in a form with one observation row per subject and each measurement as a different column, where one column identifies the treatment variable as a factor, one specifies the subgroup attribute and others contain the quantitative values for each endpoint (e.g. "Ischemic", "Hemorrhag", "Stroke") with a binary indicator depending on their manifestation, e.g. 1 for yes and 0 for no. 
New variables for every endpoint must be defined for each subgroup in order to set the other subgroup to \texttt{NA}. For example does \texttt{I.noTIA} include all data of the variable "Ischemic", only that all values of the other subgroup (patients with previous 
stroke or TIA) are set to missing.
The \texttt{mmm()} command is implemented in package \textbf{multcomp} \cite{multcomp}, which provides a framework for general linear hypotheses and multiple comparisons in parametric models. For the comparison with a control group, Apixaban vs. Aspirin, a Dunnett definition is used in the calling of \texttt{mcp()} and later passed as a list of linear functions to \texttt{mlf()}. Due to the aspect that one is interested in lower event rates in the Apixaban group and therefore positive treatment differences, the alternative hypothesis "greater" is specified.

\small{
\begin{verbatim}
library(multcomp)

# global
f1 <- glm(Ischemic ~ Treatment, data=trialdata, family=binomial())
f2 <- glm(Hemorhag ~ Treatment, data=trialdata, family=binomial())
f3 <- glm(Stroke ~ Treatment, data=trialdata, family=binomial())
# TIA (S1)
f4 <- glm(I.TIA ~ Treatment, data=trialdata, na.action = na.exclude, family = binomial)
f5 <- glm(H.TIA ~ Treatment, data=trialdata, na.action = na.exclude, family = binomial)
f6 <- glm(S.TIA ~ Treatment, data=trialdata, na.action = na.exclude, family = binomial)
# No TIA (S2)
f7 <- glm(I.noTIA ~ Treatment, data=trialdata, na.action = na.exclude, family = binomial)
f8 <- glm(H.noTIA ~ Treatment, data=trialdata, na.action = na.exclude, family = binomial)
f9 <- glm(S.noTIA ~ Treatment, data=trialdata, na.action = na.exclude, family = binomial)

# marginal MCT
fes1 <- glht(f1, mcp(Treatment = "Dunnett"), alternative="greater")
fes2 <- glht(f2, mcp(Treatment = "Dunnett"), alternative="greater")
fes3 <- glht(f3, mcp(Treatment = "Dunnett"), alternative="greater")
fes4 <- glht(f4, mcp(Treatment = "Dunnett"), alternative="greater")
fes5 <- glht(f5, mcp(Treatment = "Dunnett"), alternative="greater")
fes6 <- glht(f6, mcp(Treatment = "Dunnett"), alternative="greater")
fes7 <- glht(f7, mcp(Treatment = "Dunnett"), alternative="greater")
fes8 <- glht(f8, mcp(Treatment = "Dunnett"), alternative="greater")
fes9 <- glht(f9, mcp(Treatment = "Dunnett"), alternative="greater")

# mmm
msCI <- glht(mmm(f1,f2,f3,f4,f5,f6,f7,f8,f9), mlf(mcp(Treatment = "Dunnett")), alternative="greater")

\end{verbatim}
\normalsize

\subsection{Simultaneous analysis of the AVERROES trial}

\renewcommand{\arraystretch}{1.2}
\begin{tabular}{rr||c|cc|cc|cc}
  \hline
				&			&	effect size & \multicolumn{2}{ c| }{noadjust} & \multicolumn{2}{ c| }{Bonferroni} & \multicolumn{2}{ c }{mmm}\\
Group & Endpoint & OR & 95\% CI & $p_{}$ & 95\% CI & $p_{}$ & 95\% CI & $p_{}$ \\ 
  \hline
Global & Ischemic & 2.32 & [1.71-Inf) & 0.0000 & [1.45-Inf) & 0.0000 & [1.50-Inf) & 0.0000 \\ 
   & Hemorhag & 1.51 & [0.63-Inf) & 0.2169 & [0.40-Inf) & 1.0000 & [0.44-Inf) & 0.6519 \\ 
   & Stroke & 2.22 & [1.67-Inf) & 0.0000 & [1.43-Inf) & 0.0000 & [1.47-Inf) & 0.0000 \\ \cline{2-9}
  S1 (TIA) & Ischemic & 3.29 & [1.73-Inf) & 0.0012 & [1.22-Inf) & 0.0106 & [1.31-Inf) & 0.0077 \\ 
   & Hemorhag & 4.21 & [0.67-Inf) & 0.0999 & [0.24-Inf) & 0.8990 & [0.30-Inf) & 0.3861 \\ 
   & Stroke & 3.43 & [1.86-Inf) & 0.0004 & [1.34-Inf) & 0.0040 & [1.43-Inf) & 0.0031 \\ \cline{2-9}
  S2 (noTIA) & Ischemic & 2.09 & [1.48-Inf) & 0.0002 & [1.22-Inf) & 0.0021 & [1.27-Inf) & 0.0013 \\ 
   & Hemorhag & 1.00 & [0.35-Inf) & 0.4995 & [0.20-Inf) & 1.0000 & [0.23-Inf) & 0.9404 \\ 
   & Stroke & 1.95 & [1.41-Inf) & 0.0004 & [1.18-Inf) & 0.0035 & [1.22-Inf) & 0.0025 \\ 
   \hline
\end{tabular}
\captionof{table}{\textbf{Simultaneous inference for the AVERROES trial.} One-sided p-values and confidence intervals are computed by fitting a generalized linear model, accounting for multiplicity using Bonferroni and by the \texttt{mmm} method taking correlations into account.}


\subsection{Type I error in the simultaneous analysis of multiple subgroups}
\renewcommand{\arraystretch}{1}
\begin{tabular}{rrrrrrrrr}
  \hline
N & prop\_targ & noadjust & ttest & cellmeans & mmm & mmm.dfmax & mmm.dfmin & mmm.dfind \\ 
  \hline
20 & 0.5 & 0.0859 & 0.0458 & 0.0495 & 0.0886 & 0.0652 & 0.0436 & 0.0525 \\ 
  20 & 0.6 & 0.0800 & 0.0414 & 0.0480 & 0.0790 & 0.0584 & 0.0410 & 0.0483 \\ 
  20 & 0.7 & 0.0826 & 0.0435 & 0.0511 & 0.0812 & 0.0607 & 0.0465 & 0.0526 \\ 
  20 & 0.8 & 0.0834 & 0.0418 & 0.0517 & 0.0813 & 0.0593 & 0.0465 & 0.0523 \\ 
  50 & 0.5 & 0.0836 & 0.0443 & 0.0519 & 0.0636 & 0.0557 & 0.0478 & 0.0511 \\ 
  50 & 0.6 & 0.0817 & 0.0451 & 0.0522 & 0.0620 & 0.0553 & 0.0503 & 0.0529 \\ 
  50 & 0.7 & 0.0747 & 0.0392 & 0.0498 & 0.0582 & 0.0504 & 0.0480 & 0.0489 \\ 
  50 & 0.8 & 0.0701 & 0.0364 & 0.0488 & 0.0556 & 0.0494 & 0.0479 & 0.0489 \\ 
  100 & 0.5 & 0.0804 & 0.0456 & 0.0504 & 0.0559 & 0.0532 & 0.0496 & 0.0515 \\ 
  100 & 0.6 & 0.0803 & 0.0424 & 0.0508 & 0.0555 & 0.0522 & 0.0494 & 0.0510 \\ 
  100 & 0.7 & 0.0765 & 0.0397 & 0.0504 & 0.0566 & 0.0515 & 0.0495 & 0.0503 \\ 
  100 & 0.8 & 0.0707 & 0.0362 & 0.0508 & 0.0546 & 0.0509 & 0.0502 & 0.0506 \\ 
  500 & 0.5 & 0.0840 & 0.0425 & 0.0499 & 0.0520 & 0.0506 & 0.0496 & 0.0498 \\ 
  500 & 0.6 & 0.0750 & 0.0395 & 0.0490 & 0.0494 & 0.0487 & 0.0484 & 0.0484 \\ 
  500 & 0.7 & 0.0779 & 0.0425 & 0.0532 & 0.0542 & 0.0536 & 0.0530 & 0.0534 \\ 
  500 & 0.8 & 0.0705 & 0.0365 & 0.0501 & 0.0509 & 0.0504 & 0.0503 & 0.0504 \\ 
  1000 & 0.5 & 0.0880 & 0.0455 & 0.0521 & 0.0528 & 0.0527 & 0.0523 & 0.0525 \\ 
  1000 & 0.6 & 0.0843 & 0.0447 & 0.0544 & 0.0541 & 0.0535 & 0.0534 & 0.0534 \\ 
  1000 & 0.7 & 0.0754 & 0.0414 & 0.0523 & 0.0529 & 0.0524 & 0.0521 & 0.0521 \\ 
  1000 & 0.8 & 0.0724 & 0.0395 & 0.0529 & 0.0531 & 0.0531 & 0.0531 & 0.0531 \\ 
   \hline
\end{tabular}
\captionof{table}{\textbf{Simulated familywise error rate for simultaneous test 
procedures when testing for $H_{total}$ or $H_{target}$.} A cutout of the 
simulated familywise error rates for the hypotheses $H_{total}$ or $H_{target}$ 
at $sd=5$ with the total sample size $N$ equally distributed to the two 
treatment groups. Rates are assessed for unbalanced subgroup proportions where 
"prop\_targ" indicates the proportion of the targeted subgroup in the total population.}

\begin{tabular}{rrrrrrrrr}
  \hline
N & prop\_targ & noadjust & ttest & cellmeans & mmm & mmm.dfmax & mmm.dfmin & mmm.dfind \\ 
  \hline
20 & 0.5 & 0.1254 & 0.0455 & 0.0553 & 0.1147 & 0.0845 & 0.0415 & 0.0583 \\ 
  20 & 0.6 & 0.1184 & 0.0432 & 0.0487 & 0.1078 & 0.0763 & 0.0394 & 0.0523 \\ 
  20 & 0.7 & 0.1221 & 0.0452 & 0.0559 & 0.1176 & 0.0866 & 0.0374 & 0.0587 \\ 
  20 & 0.8 & 0.1162 & 0.0420 & 0.0486 & 0.1141 & 0.0873 & 0.0328 & 0.0543 \\ 
  50 & 0.5 & 0.1168 & 0.0430 & 0.0501 & 0.0690 & 0.0593 & 0.0471 & 0.0524 \\ 
  50 & 0.6 & 0.1199 & 0.0432 & 0.0506 & 0.0718 & 0.0606 & 0.0458 & 0.0523 \\ 
  50 & 0.7 & 0.1210 & 0.0445 & 0.0503 & 0.0746 & 0.0653 & 0.0419 & 0.0540 \\ 
  50 & 0.8 & 0.1146 & 0.0412 & 0.0532 & 0.0841 & 0.0735 & 0.0338 & 0.0524 \\ 
  100 & 0.5 & 0.1105 & 0.0387 & 0.0476 & 0.0562 & 0.0512 & 0.0456 & 0.0474 \\ 
  100 & 0.6 & 0.1129 & 0.0403 & 0.0471 & 0.0572 & 0.0522 & 0.0466 & 0.0495 \\ 
  100 & 0.7 & 0.1116 & 0.0394 & 0.0459 & 0.0560 & 0.0519 & 0.0425 & 0.0465 \\ 
  100 & 0.8 & 0.1119 & 0.0401 & 0.0488 & 0.0644 & 0.0600 & 0.0398 & 0.0501 \\ 
  500 & 0.5 & 0.1122 & 0.0400 & 0.0478 & 0.0489 & 0.0485 & 0.0476 & 0.0477 \\ 
  500 & 0.6 & 0.1147 & 0.0414 & 0.0504 & 0.0527 & 0.0516 & 0.0498 & 0.0506 \\ 
  500 & 0.7 & 0.1084 & 0.0390 & 0.0493 & 0.0500 & 0.0489 & 0.0470 & 0.0488 \\ 
  500 & 0.8 & 0.1092 & 0.0362 & 0.0480 & 0.0497 & 0.0486 & 0.0454 & 0.0471 \\ 
  1000 & 0.5 & 0.1175 & 0.0387 & 0.0478 & 0.0482 & 0.0478 & 0.0473 & 0.0476 \\ 
  1000 & 0.6 & 0.1148 & 0.0400 & 0.0485 & 0.0503 & 0.0500 & 0.0494 & 0.0493 \\ 
  1000 & 0.7 & 0.1110 & 0.0403 & 0.0506 & 0.0508 & 0.0502 & 0.0489 & 0.0491 \\ 
  1000 & 0.8 & 0.1110 & 0.0379 & 0.0487 & 0.0495 & 0.0493 & 0.0474 & 0.0489 \\ 
   \hline
\end{tabular}
\captionof{table}{\textbf{Simulated familywise error rate for simultaneous test 
procedures when testing for any hypotheses.} A cutout of the simulated 
familywise error rates for any hypotheses of ${H_{total}, H_{target}, H_{compl}}$ 
at $sd=5$ with the total sample size $N$ equally distributed to the two 
treatment groups. Rates are assessed for unbalanced subgroup proportions where 
"prop\_targ" indicates the proportion of the targeted subgroup in the total population.}

\subsection{Type I error in case of two overlapping subgroup definitions}

\begin{tabular}{rrrrrrrr}
  \hline
N & prop\_targ & noadjust & ttest & mmm & mmm.dfmax & mmm.dfmin & mmm.dfind \\ 
  \hline
20 & 0.5 & 0.1299 & 0.0434 & 0.1180 & 0.0828 & 0.0454 & 0.0586 \\ 
  20 & 0.6 & 0.1124 & 0.0393 & 0.0968 & 0.0673 & 0.0394 & 0.0485 \\ 
  20 & 0.7 & 0.1157 & 0.0412 & 0.0965 & 0.0690 & 0.0438 & 0.0536 \\ 
  20 & 0.8 & 0.1065 & 0.0413 & 0.0920 & 0.0658 & 0.0469 & 0.0537 \\ 
  50 & 0.5 & 0.1129 & 0.0410 & 0.0671 & 0.0585 & 0.0467 & 0.0508 \\ 
  50 & 0.6 & 0.1097 & 0.0409 & 0.0690 & 0.0595 & 0.0518 & 0.0546 \\ 
  50 & 0.7 & 0.0984 & 0.0384 & 0.0630 & 0.0556 & 0.0509 & 0.0527 \\ 
  50 & 0.8 & 0.0925 & 0.0312 & 0.0591 & 0.0501 & 0.0476 & 0.0484 \\ 
  100 & 0.5 & 0.1185 & 0.0416 & 0.0589 & 0.0546 & 0.0490 & 0.0516 \\ 
  100 & 0.6 & 0.1077 & 0.0379 & 0.0550 & 0.0511 & 0.0479 & 0.0490 \\ 
  100 & 0.7 & 0.0974 & 0.0366 & 0.0579 & 0.0527 & 0.0510 & 0.0513 \\ 
  100 & 0.8 & 0.0926 & 0.0321 & 0.0584 & 0.0547 & 0.0538 & 0.0538 \\ 
  500 & 0.5 & 0.1110 & 0.0406 & 0.0514 & 0.0507 & 0.0493 & 0.0497 \\ 
  500 & 0.6 & 0.1062 & 0.0384 & 0.0531 & 0.0524 & 0.0513 & 0.0515 \\ 
  500 & 0.7 & 0.0939 & 0.0362 & 0.0505 & 0.0497 & 0.0492 & 0.0493 \\ 
  500 & 0.8 & 0.0951 & 0.0358 & 0.0562 & 0.0556 & 0.0552 & 0.0553 \\ 
  1000 & 0.5 & 0.1148 & 0.0429 & 0.0533 & 0.0529 & 0.0525 & 0.0526 \\ 
  1000 & 0.6 & 0.1075 & 0.0392 & 0.0506 & 0.0505 & 0.0503 & 0.0503 \\ 
  1000 & 0.7 & 0.0994 & 0.0367 & 0.0529 & 0.0525 & 0.0525 & 0.0525 \\ 
  1000 & 0.8 & 0.0918 & 0.0338 & 0.0530 & 0.0527 & 0.0523 & 0.0524 \\ 
   \hline
\end{tabular}
\captionof{table}{\textbf{Simulated familywise error rate for simultaneous test 
procedures when testing for $H_{total}$ or $H_{target}$ in a dataset with two 
overlapping subgroups.} A cutout of the simulated familywise error rates in case 
of two overlapping subgroup definitions but one primary endpoint for the 
hypotheses $H_{total}$ or $H_{target}$ at $sd=5$ with the total sample size $N$ 
equally distributed to the two treatment groups. Rates are assessed for 
unbalanced subgroup proportions where "prop\_targ" indicates the proportion of 
the targeted subgroup in the total population.}

\begin{tabular}{rrrrrrrr}
  \hline
N & prop\_targ & noadjust & ttest & mmm & mmm.dfmax & mmm.dfmin & mmm.dfind \\ 
  \hline
20 & 0.5 & 0.1826 & 0.0437 & 0.1381 & 0.0933 & 0.0439 & 0.0599 \\ 
  20 & 0.6 & 0.1862 & 0.0421 & 0.1459 & 0.0950 & 0.0433 & 0.0615 \\ 
  20 & 0.7 & 0.1794 & 0.0437 & 0.1442 & 0.0963 & 0.0397 & 0.0615 \\ 
  20 & 0.8 & 0.1750 & 0.0425 & 0.1457 & 0.0971 & 0.0363 & 0.0611 \\ 
  50 & 0.5 & 0.1714 & 0.0423 & 0.0780 & 0.0639 & 0.0475 & 0.0537 \\ 
  50 & 0.6 & 0.1708 & 0.0387 & 0.0792 & 0.0653 & 0.0421 & 0.0520 \\ 
  50 & 0.7 & 0.1748 & 0.0443 & 0.0934 & 0.0793 & 0.0445 & 0.0603 \\ 
  50 & 0.8 & 0.1688 & 0.0395 & 0.1000 & 0.0860 & 0.0326 & 0.0546 \\ 
  100 & 0.5 & 0.1643 & 0.0386 & 0.0615 & 0.0555 & 0.0481 & 0.0503 \\ 
  100 & 0.6 & 0.1708 & 0.0399 & 0.0640 & 0.0591 & 0.0491 & 0.0524 \\ 
  100 & 0.7 & 0.1662 & 0.0366 & 0.0627 & 0.0560 & 0.0408 & 0.0479 \\ 
  100 & 0.8 & 0.1701 & 0.0384 & 0.0742 & 0.0684 & 0.0393 & 0.0507 \\ 
  500 & 0.5 & 0.1724 & 0.0399 & 0.0505 & 0.0499 & 0.0491 & 0.0493 \\ 
  500 & 0.6 & 0.1753 & 0.0435 & 0.0592 & 0.0578 & 0.0561 & 0.0567 \\ 
  500 & 0.7 & 0.1661 & 0.0412 & 0.0553 & 0.0541 & 0.0517 & 0.0533 \\ 
  500 & 0.8 & 0.1638 & 0.0352 & 0.0527 & 0.0518 & 0.0462 & 0.0483 \\ 
  1000 & 0.5 & 0.1682 & 0.0386 & 0.0484 & 0.0484 & 0.0480 & 0.0481 \\ 
  1000 & 0.6 & 0.1664 & 0.0389 & 0.0516 & 0.0506 & 0.0508 & 0.0503 \\ 
  1000 & 0.7 & 0.1714 & 0.0367 & 0.0490 & 0.0491 & 0.0472 & 0.0481 \\ 
  1000 & 0.8 & 0.1625 & 0.0369 & 0.0522 & 0.0520 & 0.0503 & 0.0513 \\ 
   \hline
\end{tabular}
\captionof{table}{\textbf{Simulated familywise error rate for simultaneous test 
procedures when testing for any hypotheses.} A cutout of the simulated 
familywise error rates in case of two overlapping subgroup definitions but one 
primary endpoint for any hypotheses of ${H_{total}, H_{target}, H_{compl}}$ 
at $sd=5$ with the total sample size $N$ equally distributed to the two 
treatment groups. Rates are assessed for unbalanced subgroup proportions where 
"prop\_targ" indicates the proportion of the targeted subgroup in the total population.}

\subsection{Type I error in the simultaneous analysis of multiple endpoints}

\begin{tabular}{rrrrrrrr}
  \hline
N & prop\_targ & noadjust & ttest & mmm & mmm.dfmax & mmm.dfmin & mmm.dfind \\ 
  \hline
20 & 0.5 & 0.1342 & 0.0373 & 0.1055 & 0.0715 & 0.0423 & 0.0560 \\ 
  20 & 0.6 & 0.1265 & 0.0332 & 0.0943 & 0.0611 & 0.0402 & 0.0513 \\ 
  20 & 0.7 & 0.1254 & 0.0365 & 0.0923 & 0.0623 & 0.0456 & 0.0550 \\ 
  20 & 0.8 & 0.1228 & 0.0367 & 0.0869 & 0.0584 & 0.0483 & 0.0546 \\ 
  50 & 0.5 & 0.1287 & 0.0380 & 0.0679 & 0.0572 & 0.0456 & 0.0510 \\ 
  50 & 0.6 & 0.1255 & 0.0354 & 0.0651 & 0.0551 & 0.0495 & 0.0523 \\ 
  50 & 0.7 & 0.1197 & 0.0333 & 0.0643 & 0.0534 & 0.0493 & 0.0515 \\ 
  50 & 0.8 & 0.1124 & 0.0323 & 0.0622 & 0.0536 & 0.0518 & 0.0528 \\ 
  100 & 0.5 & 0.1344 & 0.0337 & 0.0608 & 0.0552 & 0.0489 & 0.0514 \\ 
  100 & 0.6 & 0.1235 & 0.0351 & 0.0564 & 0.0529 & 0.0499 & 0.0516 \\ 
  100 & 0.7 & 0.1169 & 0.0324 & 0.0555 & 0.0505 & 0.0494 & 0.0505 \\ 
  100 & 0.8 & 0.1092 & 0.0270 & 0.0513 & 0.0472 & 0.0463 & 0.0464 \\ 
  500 & 0.5 & 0.1262 & 0.0372 & 0.0522 & 0.0512 & 0.0501 & 0.0506 \\ 
  500 & 0.6 & 0.1192 & 0.0337 & 0.0507 & 0.0500 & 0.0496 & 0.0499 \\ 
  500 & 0.7 & 0.1208 & 0.0350 & 0.0526 & 0.0517 & 0.0514 & 0.0515 \\ 
  500 & 0.8 & 0.1078 & 0.0293 & 0.0497 & 0.0489 & 0.0488 & 0.0489 \\ 
  1000 & 0.5 & 0.1254 & 0.0339 & 0.0485 & 0.0481 & 0.0477 & 0.0478 \\ 
  1000 & 0.6 & 0.1181 & 0.0336 & 0.0496 & 0.0490 & 0.0488 & 0.0488 \\ 
  1000 & 0.7 & 0.1171 & 0.0330 & 0.0495 & 0.0489 & 0.0487 & 0.0486 \\ 
  1000 & 0.8 & 0.1059 & 0.0273 & 0.0473 & 0.0468 & 0.0468 & 0.0469 \\ 
   \hline
\end{tabular}
\captionof{table}{\textbf{Simulated familywise error rate for simultaneous test 
procedures when testing for $H_{total}$ or $H_{target}$ in a dataset with two 
$0.8$ correlated endpoints.} A cutout of the simulated familywise error rates in case 
of two continuous endpoints with a correlation of $\rho = 0.8$ for the 
hypotheses $H_{total}$ or $H_{target}$ at $sd=5$ with the total sample size $N$ 
equally distributed to the two treatment groups. Rates are assessed for 
unbalanced subgroup proportions where "prop\_targ" indicates the proportion of 
the targeted subgroup in the total population.}

\begin{tabular}{rrrrrrrr}
  \hline
N & prop\_targ & noadjust & ttest & mmm & mmm.dfmax & mmm.dfmin & mmm.dfind \\ 
  \hline
20 & 0.5 & 0.1342 & 0.0373 & 0.1055 & 0.0715 & 0.0423 & 0.0560 \\ 
  20 & 0.6 & 0.1265 & 0.0332 & 0.0943 & 0.0611 & 0.0402 & 0.0513 \\ 
  20 & 0.7 & 0.1254 & 0.0365 & 0.0923 & 0.0623 & 0.0456 & 0.0550 \\ 
  20 & 0.8 & 0.1228 & 0.0367 & 0.0869 & 0.0584 & 0.0483 & 0.0546 \\ 
  50 & 0.5 & 0.1287 & 0.0380 & 0.0679 & 0.0572 & 0.0456 & 0.0510 \\ 
  50 & 0.6 & 0.1255 & 0.0354 & 0.0651 & 0.0551 & 0.0495 & 0.0523 \\ 
  50 & 0.7 & 0.1197 & 0.0333 & 0.0643 & 0.0534 & 0.0493 & 0.0515 \\ 
  50 & 0.8 & 0.1124 & 0.0323 & 0.0622 & 0.0536 & 0.0518 & 0.0528 \\ 
  100 & 0.5 & 0.1344 & 0.0337 & 0.0608 & 0.0552 & 0.0489 & 0.0514 \\ 
  100 & 0.6 & 0.1235 & 0.0351 & 0.0564 & 0.0529 & 0.0499 & 0.0516 \\ 
  100 & 0.7 & 0.1169 & 0.0324 & 0.0555 & 0.0505 & 0.0494 & 0.0505 \\ 
  100 & 0.8 & 0.1092 & 0.0270 & 0.0513 & 0.0472 & 0.0463 & 0.0464 \\ 
  500 & 0.5 & 0.1262 & 0.0372 & 0.0522 & 0.0512 & 0.0501 & 0.0506 \\ 
  500 & 0.6 & 0.1192 & 0.0337 & 0.0507 & 0.0500 & 0.0496 & 0.0499 \\ 
  500 & 0.7 & 0.1208 & 0.0350 & 0.0526 & 0.0517 & 0.0514 & 0.0515 \\ 
  500 & 0.8 & 0.1078 & 0.0293 & 0.0497 & 0.0489 & 0.0488 & 0.0489 \\ 
  1000 & 0.5 & 0.1254 & 0.0339 & 0.0485 & 0.0481 & 0.0477 & 0.0478 \\ 
  1000 & 0.6 & 0.1181 & 0.0336 & 0.0496 & 0.0490 & 0.0488 & 0.0488 \\ 
  1000 & 0.7 & 0.1171 & 0.0330 & 0.0495 & 0.0489 & 0.0487 & 0.0486 \\ 
  1000 & 0.8 & 0.1059 & 0.0273 & 0.0473 & 0.0468 & 0.0468 & 0.0469 \\ 
   \hline
\end{tabular}
\captionof{table}{\textbf{Simulated familywise error rate for simultaneous test 
procedures when testing for any hypotheses in a dataset with two 
$0.8$ correlated endpoints.} A cutout of the simulated familywise error rates in 
case of two continuous endpoints with a correlation of $\rho = 0.8$ for any 
hypotheses of ${H_{total}, H_{target}, H_{compl}}$ 
at $sd=5$ with the total sample size $N$ equally distributed to the two 
treatment groups. Rates are assessed for unbalanced subgroup proportions where 
"prop\_targ" indicates the proportion of the targeted subgroup in the total population.}

\footnotesize
\bibliographystyle{plain}

\end{document}